# Designing artificial zinc phosphate tribofilms with tailored mechanical properties by altering the chain length

**Authors:** Sebastian Lellig [*,1,2], Subisha Balakumar [1], Peter Schweizer [2], Eva P. Mayer [1], Simon Evertz [1], Marcus Hans [1], Damian M. Holzapfel [1], Yin Du [3], Qing Zhou [3], Martin Dienwiebel [4], Johann Michler [2,5], Jochen M. Schneider [1,6]

[*] Corresponding author

[1] Materials Chemistry, RWTH Aachen University, Aachen, Germany

[2] Laboratory for Mechanics of Materials and Nanostructures, Empa, Swiss Federal Laboratories for Materials Science and Technology, Thun, Switzerland

[3] Research & Development Institute, Northwestern Polytechnical University in Shenzhen, Shenzhen City 518063, PR China

[4] Institute for Applied Materials IAM-ZM, Karlsruhe Institute of Technology KIT, Karlsruhe, Germany

[5] EPFL - École polytechnique fédérale de Lausanne, Lausanne, Switzerland

[6] Max-Planck-Institute for Sustainable Materials GmbH, Department Structure and Nano-/Micromechanics of Materials, Group Self-Reporting Materials, Düsseldorf, Germany

**Abstract:**

Zinc dialkyldithiophosphate (ZDDP), as the most prominent lubrication additive, forms tribofilms consisting primarily of zinc phosphate glasses containing sulfides. As sulfur is linked to environmental concerns, sulfur-free zinc phosphate coatings have been sputtered from a $Zn_3(PO_4)_2$ target and investigated here.

Based on the bridging to non-bridging oxygen ratio, determined by X-ray photoelectron spectroscopy (XPS), the as deposited coatings are classified as metaphosphates. As the annealing temperature is increased, the chain lengths are reduced, as witnessed by XPS data indicated by a loss of phosphorus and oxygen of the coating surface, likely due to hydrolysis with water from the atmosphere.

Transmission electron microscopy energy-dispersive X-ray spectroscopy line scans show that the XPS-revealed composition change of the coating surface upon annealing occurs over the whole thickness of the coating. This alteration in composition and chain length reductions causes a rise in hardness, reduced Young's modulus, and wear resistance. Therefore, the properties of the artificial zinc phosphate tribofilms can be tailored via a thermally stimulated composition change, causing an alternation in chain length from meta- to orthophosphate and thereby enabling the design of coatings with desired mechanical properties.

# 1. Introduction

Regarding the environmental challenges of current times, durability and long-lastingness have turned into focus. One of the most important fields is wear reduction leading to longer life-times of highly strained components. For that, thin films are used as protective coatings [1].

Zinc dialkyldithiophosphate (ZDDP) is one of the most prominent lubricant additives used for wear reduction [2-5]. Thereby, a tribofilm is formed under load by the agglomeration of distinct islands [6-8], which consists primarily of zinc phosphate glass protecting the underlying material [2, 9]. Additionally, zinc phosphates can be used in a wide array of applications including photonics [10-12], corrosion resistance [13] and biomedical applications [14].

The zinc phosphate glasses formed by ZDDP are composed of a fundamental structure with a tetrahedral geometry, see Table 1. These orthophosphates possess a central phosphorus atom, which is surrounded by four oxygen atoms. To form longer phosphate chains, these blocks are linked via the P-O-P bond. For even longer chains, cross-linkage can occur as well [15, 16]. Chains consisting only of the basic building block are called orthophosphates. If two of these blocks link via the bridging oxygen atom, pyrophosphates are formed. Longer chains are characterized as polyphosphates, whereby infinitely long chains are labeled as metaphosphates. Ultraphosphates ($P_2O_5$) are present once cross-linkage between these chains occurs [15-17].

The chain length can be estimated by the zinc to phosphorus Zn/P ratio as zinc functions as a glass modifier, breaking up the P = O bonds and forming P-O-Zn [18]. A lower Zn/P ratio indicates longer phosphate chains [17]. To quantify the specific chain length, the bridging (BO) to non-bridging oxygen (NBO) ratio is investigated.

Hereby, the O 1s peak is split into two components, whereby the non-bridging oxygen peak is located at a binding energy 1.6 eV lower than the bridging oxygen peak. As bridging oxygen only occurs in the case that the basic building blocks are connected, the BO/NBO ratio is less dependent on the composition than the Zn/P ratio. An increase in the ratio thereby indicates an elongation of the chains [16, 17]. Characteristic values for the BO/NBO ratios are shown in Table 1.

**Table 1.** Characteristic BO/NBO ratios for different structures and chain lengths. Depictions of different chain types reconstructed after [17]. The blue, red, green, and purple dots indicate zinc, non-bridging oxygen, bridging oxygen, and phosphorus atoms, respectively.

| Structure | Chemical formula [19] | Chain length | BO/NBO [16] |
|---|---|---|---|
| orthophosphate 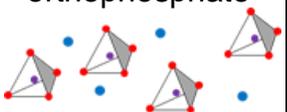 | $Zn_3(PO_4)_2$ | 1 | 0 |
| pyrophosphate 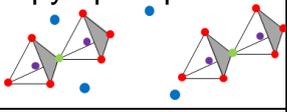 | $Zn_2P_3O_7$ | 2 | 1/6 |
| polyphosphate 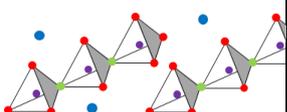 | - | > 2 | 1/6 < x > 0.5 |
| metaphosphate 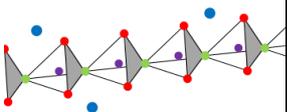 | $Zn(PO_3)_2$ | infinite | 0.5 |
| ultraphosphate 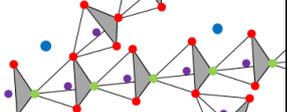 | - | cross-linked | > 0.5 |

Protective thin films formed by ZDDP can be achieved in the form of thermal films, which form at elevated temperatures, or tribofilms, which are stable in an oil environment [20], forming under mechanical loads [2, 21-23], which according to Zhang et al. is due to applied shear stresses [24]. Thereby, only in the rubbing tracks tribofilm formation can be observed, while in areas close to these tracks, the film growth is minimal [22]. While thermal films, which are formed at temperatures lower than 100 °C, consist mostly of pyrophosphates, short chains start to form at temperatures above 130 °C, which develop into longer polyphosphates close to 150 °C, which was attributed to be due to thermo-oxidative processes [2, 21]. Thermal films were shown to reach a thickness of up to 400 nm [25] with an indentation modulus of 35 GPa and a hardness of 1.5 GPa [7], whereby no iron atoms are incorporated into the thermal films for iron-containing substrates [26]. Similar compositions occur for tribofilms, even though iron phosphates are found to form within the tribofilms for iron-containing substrates [26], whereby the mechanical properties are improved. In the synthesis process, blocks of zinc phosphate glass form and merge together, keeping their general shape. Film thicknesses of up to 150 nm can be reached [8]. While most of the film consists of ortho- and pyrophosphates, a top layer of longer polyphosphates can be found [2, 7, 21, 27]. Ueda et al. [9] reported that the tribofilms can be amorphous at the surface and nanocrystalline at the interface to the substrate. This is in agreement with Nicholls et al., who showed that longer polyphosphate chains are present at the surface, while shorter chains were observed close to the substrate interface [28]. Zhang et al. reported that the tribofilm formation rate is dependent on the shear stress occurring during rubbing by reducing the thermal activation barrier for such film formation [24]. For prolonged rubbing, the films become more nanocrystalline, whereby longer phosphate chains are broken into shorter ones, which was attributed to shear stress as well as heat arising during the rubbing process, increasing their wear resistance [9]. Additionally, a correlation between the morphology of the films and the chain length was demonstrated by Canning et el., whereby larger pads in the tribofilms correlate with longer polyphosphate chains, while smaller pads possess shorter polyphosphate chains [29]. The mechanical properties of tribofilms are characterized by a hardness of 3.5 GPa and a Young's modulus of 90 GPa [2] with a typical friction coefficient of 0.115 [30].

Heuberger et al. [16] investigated zinc phosphate films formed on hardened bearing steel (100Cr6), in a ball-on-disc tribometer, under mechanical pressure with 1 wt.% ZDDP as an additive in poly-α-olefin (PAO). Hereby, a correlation between the applied pressure and the chain length as well as the thickness of the films was determined. The frictional heat generated under higher loads during the sliding process leads to longer chains and thicker films, which was attributed to thermal decomposition [21, 27]. Similarly, an increase in temperature (up to 180 °C) results in longer chain lengths and thicker tribofilms [16]. The friction coefficients of these films vary between 0.14 at room temperature to 0.2 for higher temperatures (180 °C), whereby a linear dependence was observed. This increase in friction might originate from the higher chain lengths, but could also be based on the change in film thickness [16].

Dobbalaere et al. produced zinc phosphate thin films with plasma-enhanced atomic layer deposition that were P-rich with a P/Zn ratio of 2.3, therefore consisting of metaphosphate with some ultraphosphate contributions and a maximum reported thickness of < 50 nm at a growth rate of 0.92 nm/cycle [31].

The wear resistance increase caused by ZDDP is generally attributed to the formation of phosphate-based tribolayers, hindering metal-to-metal contact in case of mixed or boundary lubrication spikes [23]. In contrast to the studies above, the goal of this work is therefore the direct synthesis of zinc phosphate glass coatings by magnetron sputtering and to demonstrate that the mechanical and wear properties can be controlled through intentional variation of the phosphate chain length. Hence, the synthesis of a sulfur-free tribolayer is demonstrated, providing an alternative to the ZDDP based, sulfur-containing tribolayers that need to be formed in the tribocontact.

To that end, magnetron sputtered zinc phosphate glass coatings were annealed, adjusting their composition and thereby their mechanical and wear properties, which is monitored by means of X-ray photoelectron spectroscopy (XPS), transmission electron microscopy (TEM), energy dispersive X-ray spectroscopy (EDX), nanoindentation and linear reciprocating wear testing.

## 2. Experimental Details

Zinc phosphate glass coatings were synthesized by radio frequency magnetron sputtering of a $Zn_3(PO_4)_2$ target (American Elements) in a high-vacuum deposition chamber on 10 x 10 mm silicon substrates. The base pressure was below $3 \cdot 10^{-7}$ mbar and sputtering was conducted in argon (99.9999 % Ar) at a pressure of 0.1 Pa with a power density of 3.45 W/cm$^2$. The deposition rate was approximately 500 nm/h at a source to substrate distance of 11 cm.

The as deposited samples were then annealed in ambient air in a GERO (SR70-200/12) tube furnace at temperatures between 100 and 400 °C in 50 °C increments. The samples were inserted into the furnace at the desired temperatures and removed after 3 hours. Additionally, samples were annealed at 400 °C in a nitrogen and oxygen atmosphere for 1 h as well as under vacuum conditions for 10 min.

To analyze the chemical composition as well as the binding characteristics of 1 µm thick coatings, XPS was performed with a Kratos AXIS Supra (AL $K_\alpha$ source: 1486.6 eV) and a pass energy of 10 eV. The chemical composition was determined from overview scans using a dwell time of 99.5 ms and a step size of 1 eV. Characteristic ratios were analyzed with detail scans, whereby 10 sweeps, a dwell time of 300 ms and a step size of 0.1 eV were averaged. A Shirley background, a mixed Gaussian-Lorentzian peak shape (30 – 70 %) and the relative sensitivity factors (RSF), provided by KRATOS, were considered. Peak calibration was performed for the C 1s peak with a binding energy of 284.8 eV.

The structures of the coatings were investigated in a Bruker AXS D8 Discover diffractometer, including a General Area Diffraction System (GADDS) using a copper target (Cu K$_\alpha$: λ = 1.5418 Å) with X-ray diffraction (XRD). Hereby, 2θ was varied between 30 and 90 ° with an accelerating voltage of 40 V and a current of 40 mA.

For the in-depth characterization, as well as the analysis over the whole coating thickness, STEM imaging and EDX mapping was conducted in a TEM Themis 200 G3 with a SuperX detector. A FEI Helios Nanolab 660 dual-beam microscope was used to prepare the TEM lamella by focused ion beam (FIB) milling with gallium ions and the subsequent milling to a thickness of around 80 nm.

To analyze the mechanical properties of the coatings, nanoindentation measurements were conducted with a Hysitron TI-900 TriboIndenter using the Oliver-Pharr method [32]. A fused silica standard was used to calculate the tip area function of the Berkovich diamond tip. With a load of 200 µN for the 5x5 indents per sample, a maximum contact depth of 35 nm for the coating thickness of ~ 2 µm was reached.

Wear resistance was investigated using a reciprocating linear tribometer Tetra Basalt MUST. The tests were done at room temperature with a 100Cr6 bearing steel ball with a 10 mm diameter as a counterpart under dry conditions as well as with a 9.65 mm diameter under poly-α-olefin (PAO) lubrication. Tests were performed at a sliding speed of 0.5 mm/s for 100 cycles, a stroke length of 1.5 and 3 mm and a load of 1 and 5 N, resulting in contact pressures of ~ 180 MPa and ~ 402 MPa for dry and lubricated conditions, respectively.

## 3. Results and Discussion

**Chemical Analysis**

Zinc phosphate glass coatings deposited from a $Zn_3(PO_4)_2$ target at room temperature possess an X-ray amorphous structure, see supplementary figure S1. XPS analysis reveals a chemical composition of approximately 30 at.% Zn, 23 at.% P, 38 at.% O and 9 at.% C on the surface. Additionally, the ratio of the bridging (green curve) and non-bridging oxygen (red curve) peaks constituting the O 1s peak, was determined (see Figure 1). According to Brow et al. [15] and Crobu et al. [17], the phosphate chain length can be determined based on the BO/NBO of the O 1s signal. As shown in Figure 1, phosphate glass is composed of tetrahedral building blocks, where bridging oxygen, colored in green, is connecting the individual phosphate building blocks, which contain non-bridging oxygen, colored in red. Therefore, an increasing BO/NBO ratio corresponds to an increase in phosphate chain length, see Table 1 and Figure 1. Hence, the shortest possible chain is characterized by the absence of bridging oxygen with a corresponding BO/NBO ratio of 0, see "orthophosphate" [16, 17], Figure 1. If two of the tetrahedral building blocks are connected via one bridging oxygen atom, the BO/NBO ratio of the so formed pyrophosphate is 1/6 [16, 17], see "pyrophosphate", Figure 1. The more tetrahedra are connected via bridging oxygen atoms, the longer the chains are, whereby chains within the BO/NBO ratio regime between 1/6 and 0.5 are characterized by chain lengths > 2, see "polyphosphate", Figure 1. Once all building blocks are connected (without cross-links), metaphosphate chains form with a BO/NBO ratio of 0.5, see "metaphosphate" [16, 17], Figure 1.

Considering the above discussion, a qualitative analysis of the XPS data of the surface shown in Figure 1, indicates a decrease in chain length as the annealing temperature is increased. In Figure 2, the BO/NBO ratio of the surface is plotted against the annealing temperature to enable a quantitative analysis of thermally induced changes in the phosphate chain length. In the as deposited state, metaphosphate chains seem to dominate, correlating to a BO/NBO ratio of 0.5 [16]. After annealing at 100 °C, metaphosphates that have started to cross-link (ultraphosphates), are present, while in the temperature region from 100 to 250 °C, the BO/NBO ratio decreases linearly from > 0.5 to 1/6, implying a transition towards pyrophosphates [21]. After annealing in the temperature range of 300 to 400 °C, the formation of orthophosphates with a BO/NBO ratio of 0 is observed.

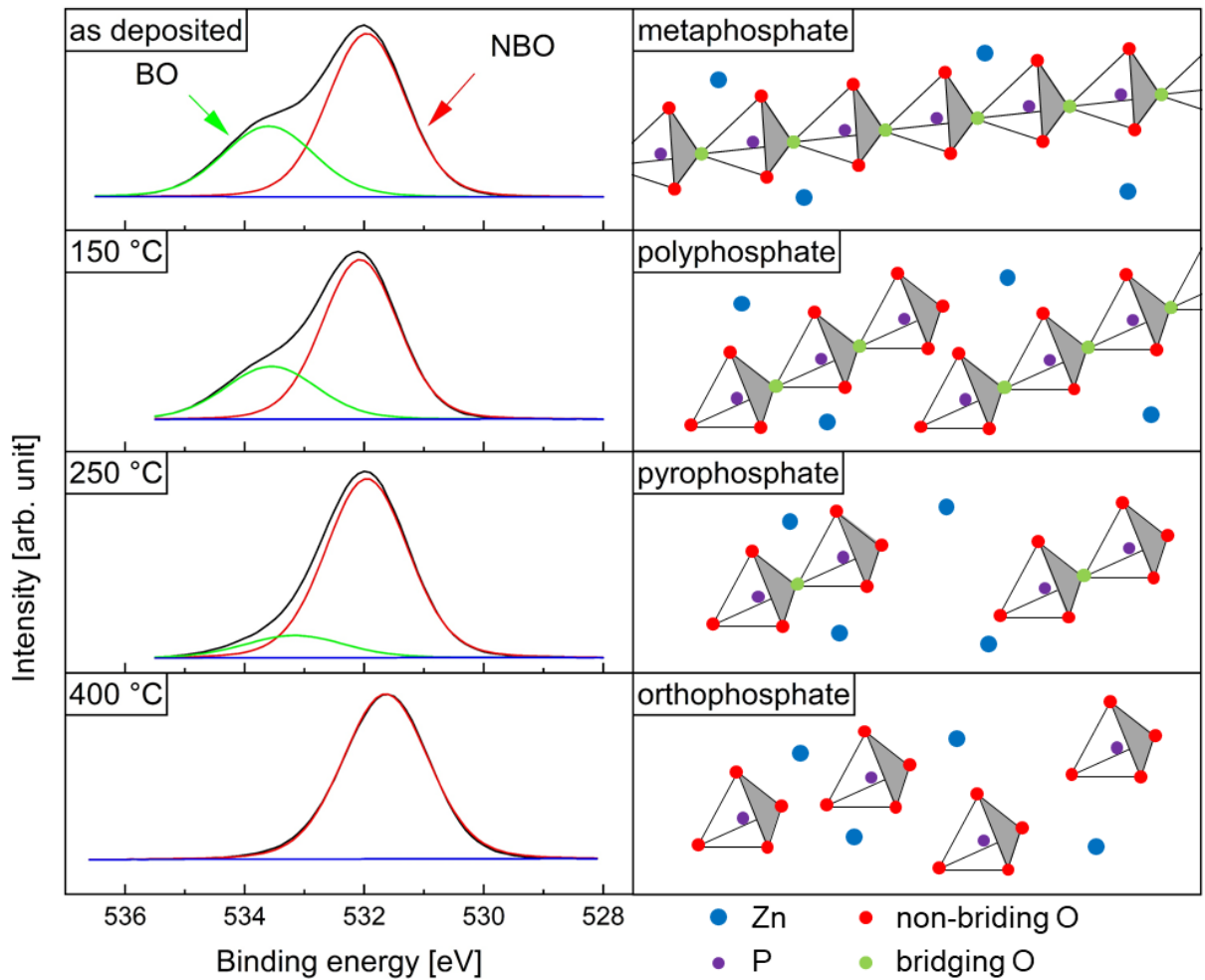

**Figure 1.** XPS O 1s peak split into bridging (BO) and non-bridging oxygen (NBO) for meta-, poly-, pyro- and orthophosphate. With decreasing BO content, the chain length is decreased, which correlates with the annealing temperature [16]. Depictions of different chain types reconstructed after [17].

Heuberger et al. [16] investigated the formation of tribofilms in the tribological contact in the presence of ZDDP containing PAO lubricant at temperatures ranging from room temperature to 180 °C. The hereby obtained BO/NBO ratios were in the range of 0.1 to 1, indicating that tribofilms with similar chain lengths as obtained here are formed. Thereby, an initial alkylation of S in ZDDP is followed by a reaction of neighboring phosphoryl groups, which leads to the formation of longer poly- and cross-linked sulfur-containing phosphates [16]. In contrast, the here directly synthesized artificial sulfur-free, metaphosphate tribofilms can be transformed via thermally stimulated composition changes into phosphates with shorter chains, ultimately resulting in the formation of orthophosphates.

Therefore, sputter deposition with subsequent annealing was successfully used to reach a similar coating composition and chain lengths as in sulfur-containing tribofilms, generated in a tribocontact of hardened bearing steel 100Cr6, when lubricated by a ZDDP containing oil [16]. Hence, artificial zinc phosphate coatings can be designed with a desired chain length.

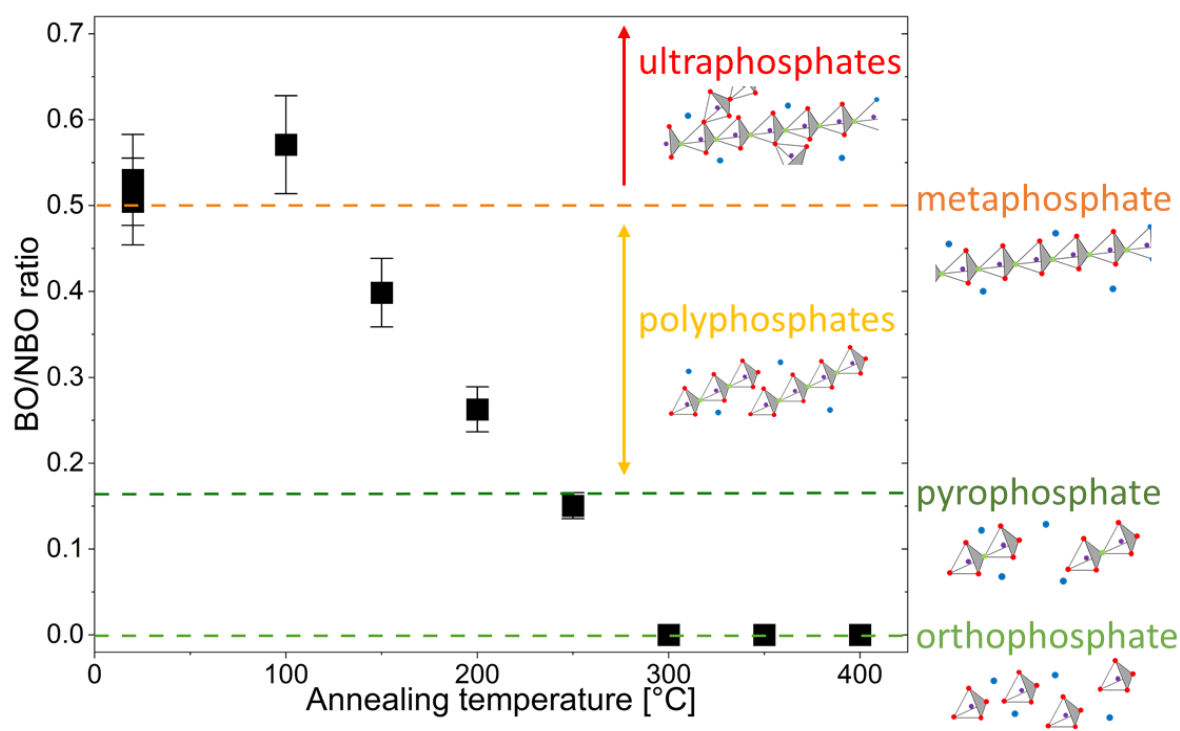

**Figure 2.** Characteristic BO/NBO ratio obtained by XPS is used to determine the chain length. With increasing annealing temperature, a decrease in the chain length is observable. By sputtering and post-annealing, the same range of phosphates as reported by Heuberger et al. [16] can be achieved in a controlled manner with a linear decrease in chain length. Depiction of phosphate types modified after [17], classification into phosphate types based on BO/NBO ratios [16].

To investigate, whether the composition alterations in the surface due to annealing, as probed by XPS, are also occurring in the bulk of the coating, EDX line scans were conducted in TEM for the as deposited sample, see also supplementary figure S2, as well as for a sample after annealing at 400 °C. As can be seen in Figure 3, a decrease in oxygen and phosphorus content relative to the zinc content can be observed over the whole thickness of the coating. This is in agreement with the XPS data and it can be concluded, that a homogeneous alteration of composition and, therefore, chain length over the whole coating was achieved with the above-described synthesis method.

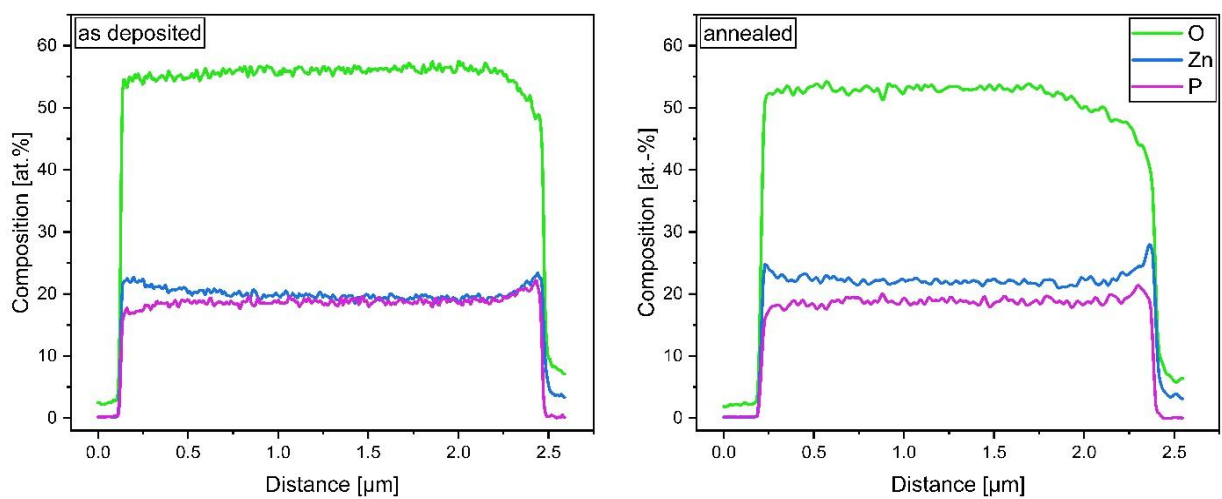

**Figure 3**. TEM EDX linescan of as deposited sample and annealed at 400 °C for 3 h. The reduction of phosphorus in comparison to zinc is visible over the whole coating.

**Proposed mechanism**

To gain a deeper understanding of the mechanism that leads to the formation of shorter chains, samples were annealed at 400 °C in different atmospheres in the XPS load-lock and analyzed directly with XPS. In Figure 4 a), the BO/NBO ratios for coatings annealed in nitrogen, vacuum and oxygen are shown. While for all samples, a small decrease in BO/NBO and therefore a reduction in chain length is observable, the magnitude is significantly smaller than for samples annealed in air. This may be explained by the significantly smaller water partial pressure in nitrogen, vacuum and oxygen compared to ambient air, where the relative humidity was approximately 78 % [33]. In both the XPS and EDX data, see supplementary figure S3 and Figure 3, a reduction of phosphorus and oxygen in relation to zinc was observable, with a P/Zn and O/Zn ratio of 0.79 and 1.28 for the as deposited coating, decreasing to 0.32 and 0.59 after annealing in ambient air at 400 °C, respectively. Zinc phosphate glasses are known to be prone to hydrolysis, whereby the phosphate chains, consistent with the observations reported here, are broken [34, 35] via a reaction with water and by protonation, phosphoric acid is formed (see Figure 4 b)) [36].

Thereby, water molecules diffuse into the bulk of the coating, initiating hydrolysis in addition to the surface [35]. Furthermore, the surplus of zinc, compared to the as deposited state, that is created during annealing amplifies the effect as zinc functions as a glass modifier, breaking up the chains via reactions with bridging oxygen [19].

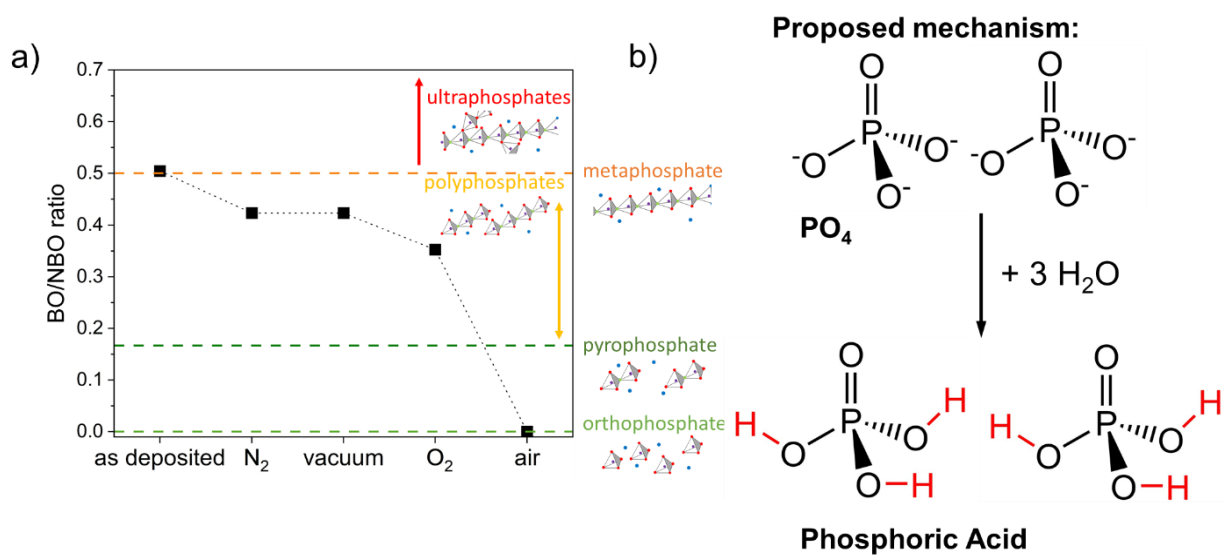

**Figure 4.** a) BO/NBO ratio for samples annealed at 400 °C in nitrogen, vacuum and oxygen. All of them show a slight decrease in chain length, but not as pronounced as in air. The line connecting the datapoints is a guide to the eye. b) Proposed mechanism of hydrolysis due to the moisture in air is presented as a possible reason for this deviation.

**Influence of chain length alteration on mechanical properties**

To evaluate the influence of the chain length modification on the mechanical behavior, nanoindentation was performed using ~ 2 µm thick coatings and the annealing protocol from Figure 2. Assuming that the annealing-induced changes in chain length are independent of coating thickness, hardness, see Figure 5 a), and reduced Young's modulus values, see Figure 5 b), were plotted versus the BO/NBO ratio determined for the 1 µm thick coatings discussed in Figure 2. Figure 5 a) shows the hardness as a function of the BO/NBO ratio. In general, a trend to lower hardness values for longer chain lengths can be observed. The hardness values obtained for longer chains with BO/NBO ratios ranging from 0.57 to 0.26 are less than 4 GPa, similar to the reported hardness values of 3 and 4 GPa for ZDDP tribofilms with BO/NBO ratios of 0.54 and 0.27, respectively [37]. In contrast, coatings with shorter chain lengths possess hardness values of up to 5.5 GPa at a BO/NBO ratio of 0. A similar trend is visible in Figure 5 b) for the reduced Young's modulus as a function of BO/NBO ratio, with an overall decrease in reduced Young's modulus from 84 to 65 GPa for longer chain lengths.

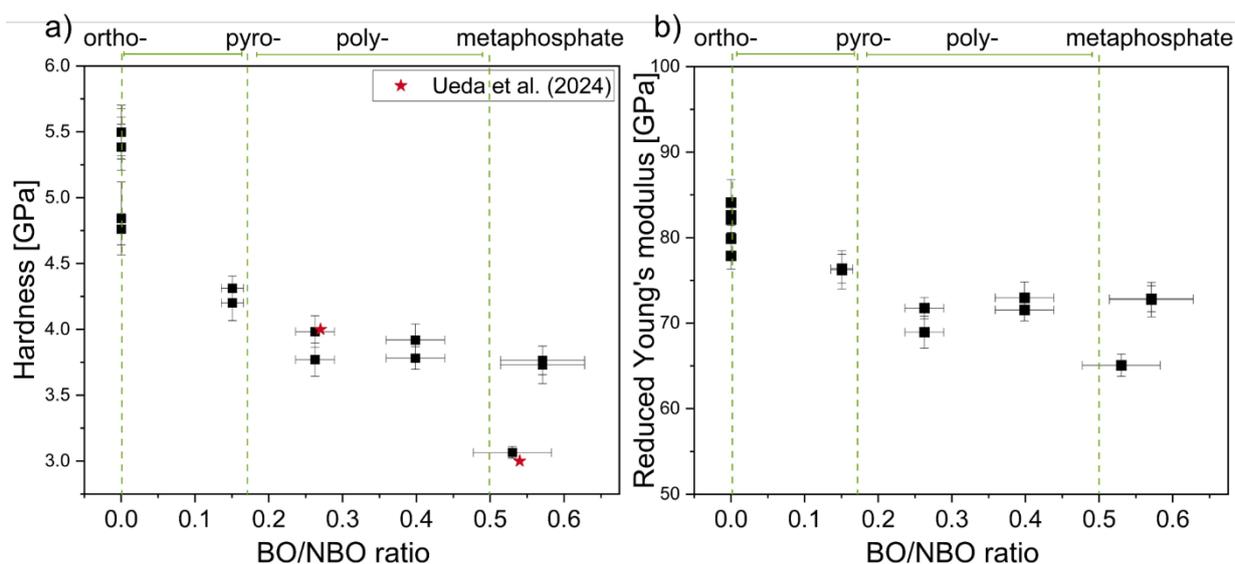

**Figure 5.** Hardness and reduced Young's modulus for different chain lengths from annealed samples as well as hardness values for different BO/NBO ratios obtained from Ueda et al. [37] for tribofilms generated in a ball-and-disc tribometer on AISI52100 in a ZDDP solution. With decreasing chain length, both hardness and Young's modulus are increased up to the orthophosphate chain length.

**Wear and friction behavior**

To probe the influence of the chain length on the tribological behavior, the as deposited coating as well as a sample annealed at 400 °C were analyzed in a reciprocating linear tribometer. The results shown in Figure 6 a) demonstrate a comparable friction coefficient for the annealed and as-deposited sample of ~ 0.1 in a PAO lubrication environment. A steady state friction coefficient of ~ 0.12 was reached after 400 s of run-in, whereby the annealed sample exhibits with 0.017 a factor 6 lower initial friction coefficient than the as deposited coating. This might be due to differences in surface roughness of the annealed sample composed of orthophosphate as well as the higher hardness and reduced Young's modulus in comparison to the metaphosphate of the as deposited sample. The friction coefficients are comparable to tribofilms formed by ZDDP as reported by Heuberger et al. (between 0.14 and 0.2 for identical load and increasing temperature) [16]. The increase in friction coefficient for increasing temperatures in the case of Heuberger is most probably determined by larger film thicknesses and not due to the change in chain length.

The wear depths of the as deposited coating and the at 400 °C annealed sample are shown in Figure 6 b). The maximal wear depth of the annealed sample is 54 % smaller than the one measured for the as-deposited sample. Thereby, while the steady state friction coefficient is similar for both coatings, the post-deposition annealing treatment leads not only to a strongly reduced phosphate chain length, but also to a significantly increased wear resistance of the protective coating, reducing the wear volume of the as deposited sample of $3.0 \cdot 10^{14}$ nm$^3$ by factor 3.5 to a wear volume of $8.6 \cdot 10^{13}$ nm$^3$ for the annealed sample, correlating with the higher hardness and reduced Young's modulus for the annealed sample, see Figure 5. This is in agreement with Ueda et al. who reported an increased wear resistance for ZDDP tribofilms with shorter phosphate chains forming upon longer rubbing [9].

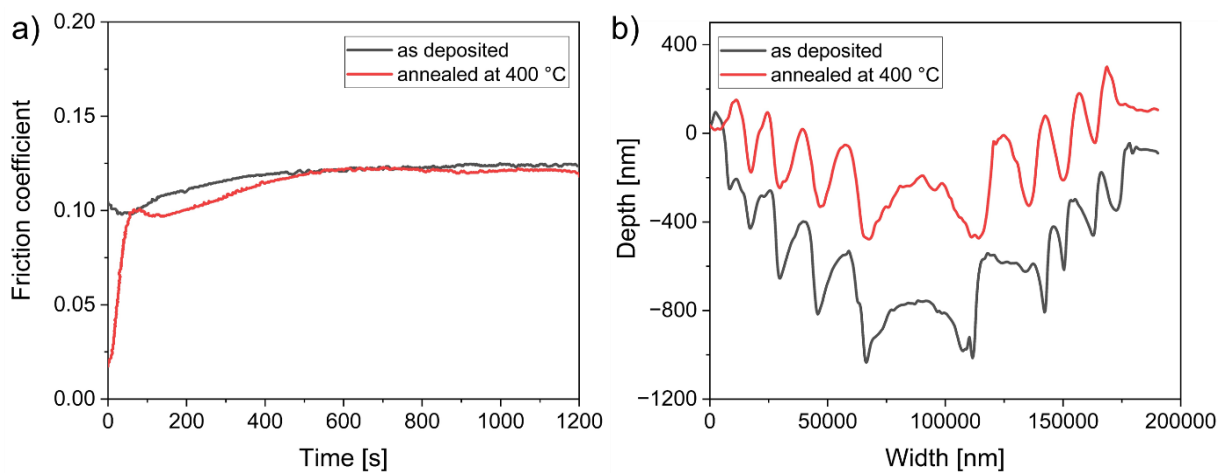

**Figure 6.** Friction coefficient in PAO lubrication environment and wear track depths in dry environment after 100 cycles and a load of 5 N. Wear resistance of the annealed sample (400 °C) is increased in comparison to as deposited sample.

## 4. Summary

Annealing-induced composition modifications of magnetron sputtered sulfur-free zinc phosphate glass coatings were shown to systematically alter the chain lengths from meta- to orthophosphates, likely by hydrolysis with water in the atmosphere, enabling the controlled tailoring of mechanical properties. Thereby, artificial tribofilms with similar chain lengths as were obtained for sulfur-containing tribofilms, generated in a ZDDP-containing oil in a hardened bearing steel 100Cr6 tribocontact, were synthesized directly. Overall, the decrease in chain length as probed by XPS can be attributed to a loss of phosphorus and oxygen upon annealing. Simultaneously, nanoindentation revealed that the change in chain length and composition linearly influenced the mechanical properties, increasing the hardness and reduced Young's modulus by 79 % and 29 %, respectively, providing a useful tool to design coatings with desired properties. Thereby, also the wear resistance for samples with shorter chain lengths is increased. In summary, artificial zinc phosphate glass tribofilms were synthesized that allow precise designing of the mechanical and wear properties to desired values by altering the chain length in a post-annealing treatment.

## 5. Acknowledgments

The authors would like to acknowledge the Guangdong Basic and Applied Basic Research Foundation (2024A1515012378).

## 6. Supplementary Information

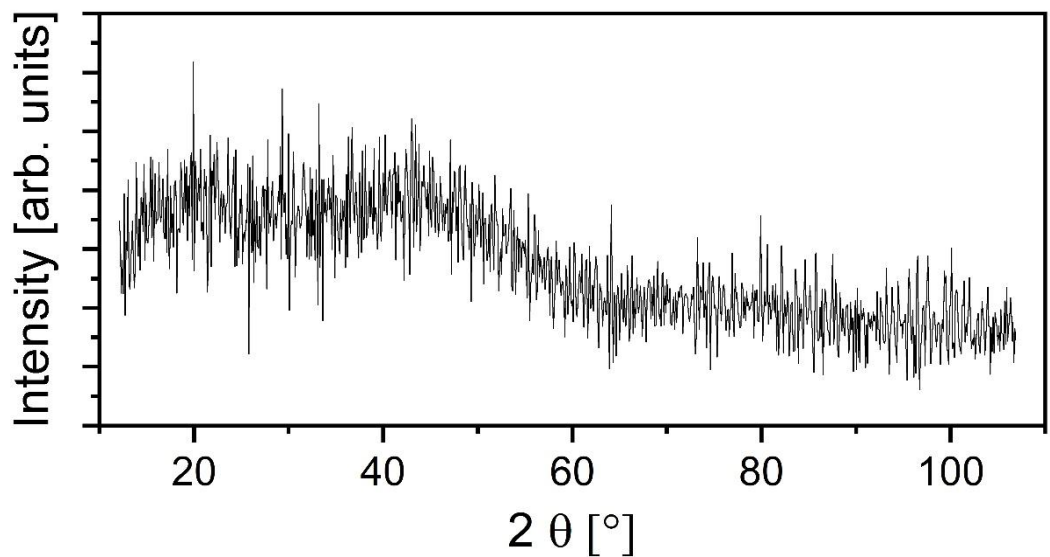

**Figure S 1.** XRD of as deposited zinc phosphate sample shows an amorphous structure.

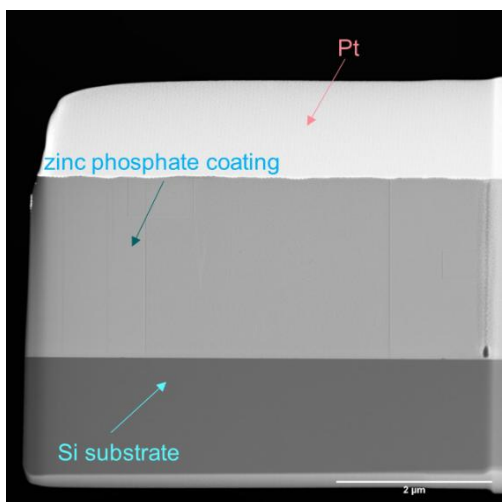

**Figure S 2.** STEM image of as deposited zinc phosphate glass coating.

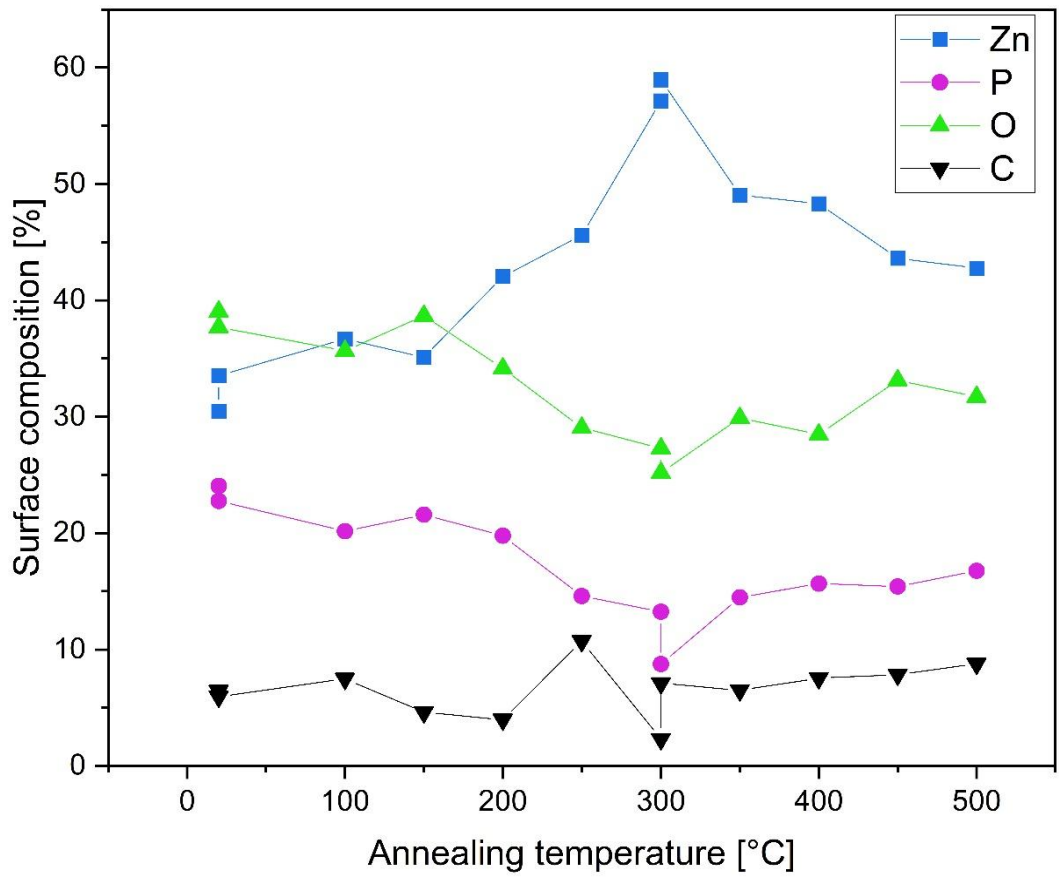

**Figure S 3.** Surface composition, obtained by XPS, for zinc phosphate coatings upon annealing.